\documentclass[sn-mathphys-num]{sn-jnl}

\usepackage{graphicx}%
\graphicspath{{./Figs/}}

\usepackage{multirow}%
\usepackage{amsmath,amssymb,amsfonts}%
\usepackage{amsthm}%
\usepackage{mathrsfs}%
\usepackage[title]{appendix}%
\usepackage{xcolor}%
\usepackage{textcomp}%
\usepackage{manyfoot}%
\usepackage{booktabs}%
\usepackage{algorithm}%
\usepackage{algorithmicx}%
\usepackage{algpseudocode}%
\usepackage{listings}%
\usepackage{geometry}
\usepackage{ulem}
\usepackage{hyperref}

\theoremstyle{thmstyleone}%
\theoremstyle{thmstyletwo}%

\theoremstyle{thmstylethree}%

\begin{document}

\title[Gromov-Wasserstein unsupervised alignment reveals structural correspondences between the color similarity structures of humans and large language models]{Gromov-Wasserstein unsupervised alignment reveals structural correspondences between the color similarity structures of humans and large language models}


%
\author*[1]{\fnm{Genji} \sur{Kawakita}}\email{g.kawakita22@imperial.ac.uk}

\author[2,3]{\fnm{Ariel} \sur{Zeleznikow-Johnston}}\email{ariel.zeleznikow-johnston@monash.edu}

\author[2,3,4,5]{\fnm{Naotsugu} \sur{Tsuchiya}}\email{naotsugu.tsuchiya@monash.edu}

\author*[6]{\fnm{Masafumi} \sur{Oizumi}}\email{c-oizumi@g.ecc.u-tokyo.ac.jp}

\affil[1]{\orgdiv{Department of Bioengineering}, \orgname{Imperial College London}, \orgaddress{\city{London}, \country{UK}}}

\affil[2]{\orgdiv{School of Psychological Sciences}, \orgname{Monash University}, \orgaddress{\city{Melbourne}, \country{Australia}}}

\affil[3]{\orgdiv{Turner Institute for Brain and Mental Health}, \orgname{Monash University}, \orgaddress{\city{Melbourne}, \country{Australia}}}

\affil[4]{\orgdiv{Center for Information and Neural Networks (CiNet)}, \orgname{National Institute of Information and Communications Technology (NICT)}, \orgaddress{\city{Osaka}, \country{Japan}}}

\affil[5]{\orgdiv{Department of Qualia Structure, ATR Computational Neuroscience Laboratories}, \orgaddress{\city{Kyoto}, \country{Japan}}}

\affil[6]{\orgdiv{Graduate School of Arts and Science}, \orgname{The University of Tokyo}, \orgaddress{\city{Tokyo}, \country{Japan}}}


\abstract{Large Language Models (LLMs), such as the General Pre-trained Transformer (GPT), have shown remarkable performance in various cognitive tasks. However, it remains unclear whether these models have the ability to accurately infer human perceptual representations. Previous research has addressed this question by quantifying correlations between similarity response patterns of humans and LLMs. Correlation provides a measure of similarity, but it relies pre-defined item labels and does not distinguish category- and item- level similarity, falling short of characterizing detailed structural correspondence between humans and LLMs. To assess their structural equivalence in more detail, we propose the use of an unsupervised alignment method based on Gromov-Wasserstein optimal transport (GWOT). GWOT allows for the comparison of similarity structures without relying on pre-defined label correspondences and can reveal fine-grained structural similarities and differences that may not be detected by simple correlation analysis. Using a large dataset of similarity judgments of 93 colors, we compared the color similarity structures of humans (color-neurotypical and color-atypical participants) and two GPT models (GPT-3.5 and GPT-4). Our results show that the similarity structure of color-neurotypical participants can be remarkably well aligned with that of GPT-4 and, to a lesser extent, to that of GPT-3.5.These results contribute to the methodological advancements of comparing LLMs with human perception, and highlight the potential of unsupervised alignment methods to reveal detailed structural correspondences. This work has been published in \textit{Scientific Reports}, DOI: \href{https://doi.org/10.1038/s41598-024-65604-1}{10.1038/s41598-024-65604-1}.}


\maketitle

Large Language Models (LLMs) have demonstrated remarkable performance in a variety of cognitive tasks \cite{Devlin2018, Bubeck2023, Binz2023}. These LLMs, based on the Transformer architecture, use self-attention mechanisms to effectively process and generate sequences of data \cite{Vaswani2017}. Among LLMs, the General Pre-trained Transformer (GPT) series developed by OpenAI has received considerable public attention with the introduction of the ChatGPT conversational interface \cite{Brown2020, OpenAI2023}. Recently introduced GPT models can generate human-like responses to prompts, and are reported to excel at tasks assessing Theory of Mind abilities \cite{Kosinski2023}.

These observations raise two intriguing questions: To what extent can Large Language Models accurately infer human perceptual representations, and how can we effectively compare LLMs and human perceptual representations? Previous research addressed this by comparing human similarity judgments with those generated by GPT across various modalities \cite{Marjieh2023}. They found correlations as high as $\rho=0.8$ between human and LLM color similarity judgments.

To evaluate the similarity between representational structures, a supervised approach known as Representational Similarity Analysis (RSA) has been widely used in neuroscience to compare different similarity matrices obtained from behavioral, neural, and neural network model data \cite{Kriegeskorte2013, Roads2024}. The supervised approach (or supervised alignment method in general \cite{Williams2021}) assumes that an element in one similarity structure (for example, the color 'red') corresponds to the same element in another similarity structure. It then quantifies the degree of similarity between the different structures, assuming a one-to-one correspondence as defined by the item labels. Previous studies comparing the similarity structures of humans and LLMs also use this supervised approach \cite{Marjieh2023, Marjieh2022a, Marjieh2022b}.

While high correlations suggest remarkable representational similarity between humans and LLMs, interpreting their significance is challenging. First, correlation values lack appropriate controls, such as simple color space models (e.g., RGB or LAB), for comparison. These controls serve as baselines to determine whether the high correlations between human and LLM similarity judgments are truly indicative of sophisticated representational similarity or if they can be achieved by simpler models. Second, high correlations may indicate only coarse category-level alignment without capturing fine-grained structural correspondence \cite{Kriegeskorte2013, Sasaki2023}.

Regarding the first point, simple correlation can only be interpreted relative to other representational models, especially because simple correlation does not provide an absolute measure of representational equivalence. For example, even if correlation values in the range of 0.7-0.8 seem impressively high in the context of a color similarity judgment task, this does not necessarily mean that such values can only be achieved by sophisticated neural network models such as GPT. If simpler color space models, such as RGB or LAB, can achieve similar correlation levels with human judgments, the significance of a high human-to-GPT correlation becomes less pronounced. 

Regarding the second point, simple correlation does not necessarily entail a fine-item-level structural correspondence between two similarity structures. For example, previous studies using representational similarity analysis (RSA) \cite{Kriegeskorte2013}, a common method for assessing perceptual representational similarity via simple correlations between similarity matrices, have shown that high correlation values (e.g., $\rho=0.95$) may indicate only coarse category-level correspondence, even while fine-item-level alignment is completely absent (e.g., Fig. 3 in \cite{Sasaki2023}). Thus, the mere presence of a high correlation does not clarify whether the structures have a fine-item-level alignment or simply a coarse-category-level correspondence.

\begin{figure}[t]
\centering
\includegraphics[width=0.9\textwidth]{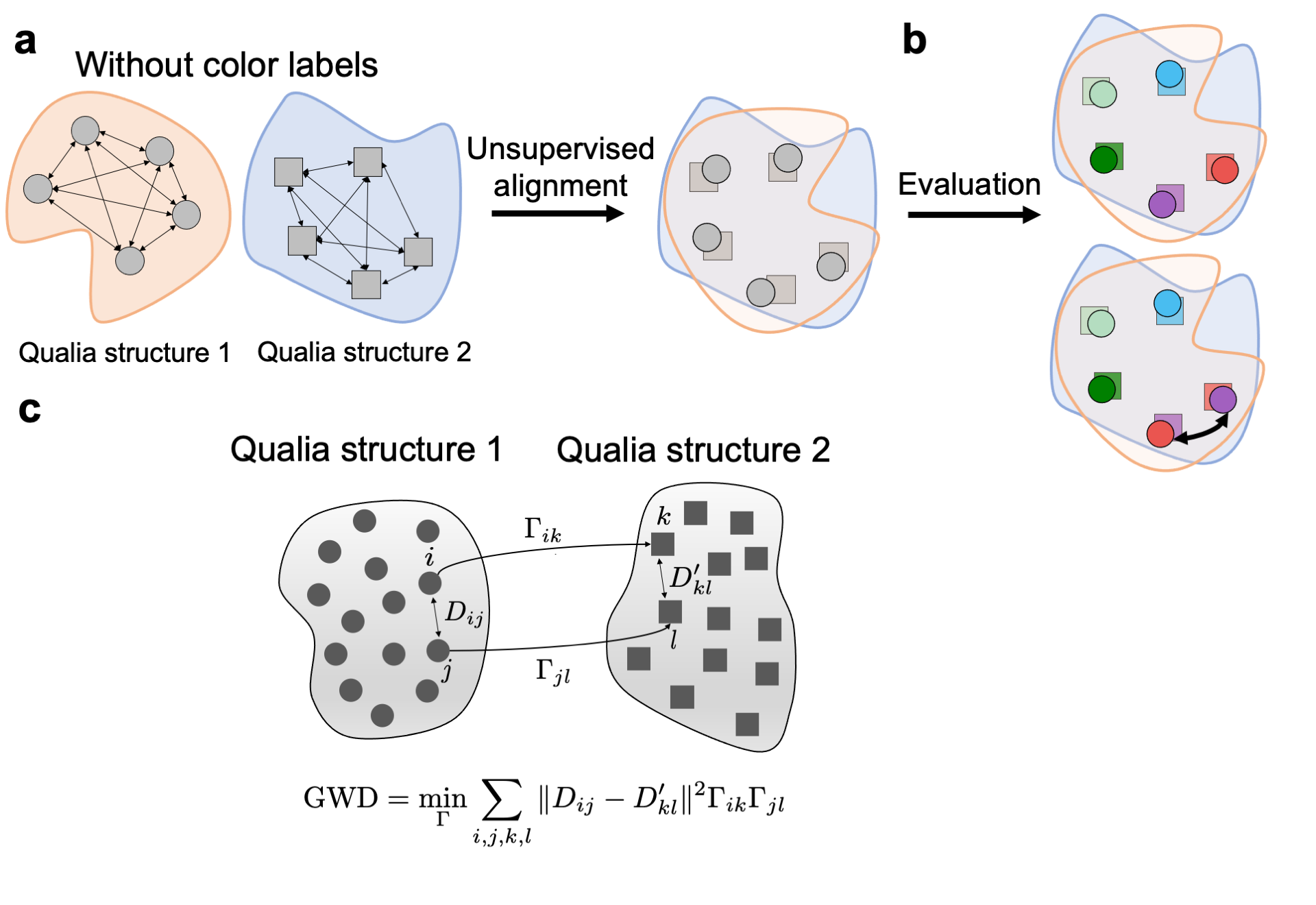}
\caption{\textbf{Schematic of unsupervised alignment.} 
(a) Unsupervised alignment of similarity structures without external labels, based only on similarity relations. (b) Evaluation of unsupervised alignment using external labels. (c) Schematic of Gromov-Wasserstein optimal transport. The elements of matrices $D$ and $D'$ are the dissimilarities between the items. $\Gamma$ is the transportation matrix, where each element indicates the probability of an item in one similarity structure corresponding to another item in the other similarity structure. Modified from \cite{Kawakita2023}. 
}
\label{unsupervised}
\end{figure}

To address these limitations, we propose using an unsupervised alignment approach to assess a more detailed level of structural correspondence between the similarity structures of humans and LLMs. In unsupervised alignment, the correspondence between items in two similarity structures is not assumed. Instead, the correspondences need to be discovered through an alignment procedure (Fig. \ref{unsupervised}a), since information about external item labels is not used. After alignment, external labels are used only to evaluate the alignment (Fig. \ref{unsupervised}b).

For unsupervised alignment, Gromov-Wasserstein optimal transport (GWOT) \cite{Memoli2011} (Fig. \ref{unsupervised}c) has recently emerged as a promising method for comparing and aligning similarity structures. GWOT has been successfully applied in various contexts, such as aligning word embedding spaces across languages \cite{Alvarez-Melis2018}, single-cell multi-omics data \cite{Demetci2020}. The unsupervised optimal transport method has revealed the structural correspondence of the similarity structures of colors across individuals \cite{Kawakita2023} and objects \cite{Sasaki2023}, facilitating a broad structural exploration of human perceptual structures. These advances in unsupervised alignment techniques provide new ways to understand the extent to which LLMs can accurately infer human perceptual structures.

Using GWOT, we compared the color similarity structures of color-neurotypical and color-atypical human participants with those of GPT-3.5, GPT-4, and color space models. Our larger 93-color dataset, compared to the 23 colors in \cite{Marjieh2023}, allows studying higher-dimensional color similarity. We also contrasted GPT-4 and GPT-3.5 to explore the effects of visual input integration and model/data size. As baselines, we considered color-atypical individuals and simple color space models (RGB and LAB). The inclusion of RGB and LAB is important to rule out the possibility that GPT responses are based solely on these models.

Our results show that the color similarity structure of color-neurotypical participants can be remarkably well aligned with that of GPT-4 and, to a lesser extent, with that of GPT-3.5. In contrast, the color similarity structures of color-neurotypical participants could not be aligned with those of color space models, despite reasonably high correlation values. These findings suggest a strong fine-item-level structural correspondence between color-neurotypical human participants and the recent GPT models, but not between color-neurotypical human participants and color-space models. The results provide insights into LLMs' ability to capture human color perception and demonstrate the utility of unsupervised alignment methods in revealing detailed structural similarities and differences between human and LLM representations.

\section*{Results}
\begin{figure}[t]
\centering
\includegraphics[width=0.9\textwidth]{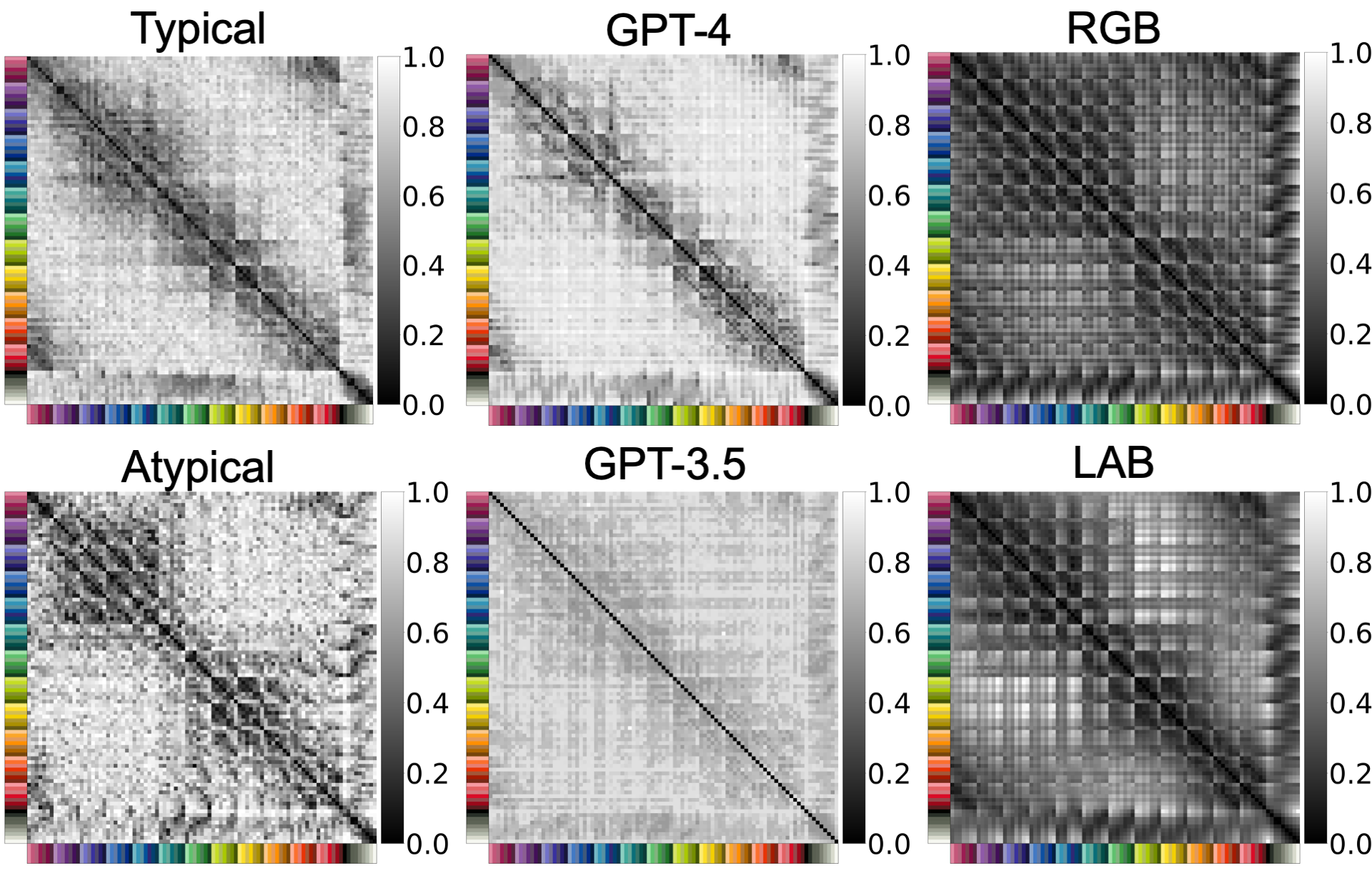}
\caption{\textbf{Color dissimilarity matrices.} Displayed are the dissimilarity matrices of 93 colors from the color-neurotypical participants group, the color-atypical participants group, GPT-4, GPT-3.5, and the RGB and LAB color space models. All matrices are normalized to have values between 0 and 1, where 0 means the no difference between colors and 1 means the maximum difference for each dissimilarity matrix.
}\label{Fig:sim}
\end{figure}

\subsection*{Color dissimilarity matrices}
In this study, we compared the similarity structures of 93 colors obtained from human participants (color-neurotypical and color-atypical participants) with those of large language models (GPT-3.5 and GPT-4). As for the human participants data, we used a large-scale dataset including 426 color-neurotypical participants and 257 color-atypical participants, which we previously collected \cite{Kawakita2023}. Color-atypical participants in this study refer to individuals with red-green color blindness, who were screened using a modified online Ishihara test (see supplementary material for details on inclusion criteria and screening procedure). In the color similarity judgement experiment, human participants were asked to rate the perceived similarity between pairs of colors drawn from a set of 93 color stimuli. For each pair, participants provided a rating on a scale from 0 to 7, with 0 indicating that the colors were perceived as very similar and 7 indicating that the colors were perceived as very different. The details of the experimental design and procedure can be found in the supplementary information. We obtained the dissimilarity matrices of 93 colors for the color-neurotypical and color-atypical participants group by simply averaging the similarity judgement responses for each color pair from all the participants in each participant group as shown in Fig. \ref{Fig:sim}. 

To obtain responses from GPT-3.5 (\verb|gpt-3.5-turbo|) and GPT-4 (\verb|gpt-4-0314|), we used a prompt that represented each color as a HEX code in line with a previous study \cite{Marjieh2023} (see Methods for the details). By collecting a complete set of similarity judgements of 93 colors with the same prompt, we obtained the dissimilarity matrices of GPT-3.5 and GPT-4 as shown in Fig. \ref{Fig:sim}. By visual inspection, we can clearly see that the dissimilarity matrix of GPT-4 is very similar to that of the color-neurotypical participants. 

To examine the possibility that LLMs rely on established color space models when judging color similarity, we computed the dissimilarity matrices for the 93 colors using two representative color space models, RGB and LAB color space models, as shown in Fig. \ref{Fig:sim} (see Methods for the details). These simple color space models served as baselines to evaluate how well large language models can approximate the human color similarity structures.

\subsection*{Correlations between color dissimilarity matrices}
\begin{figure}[t]
\centering
\includegraphics[width=0.9\textwidth]{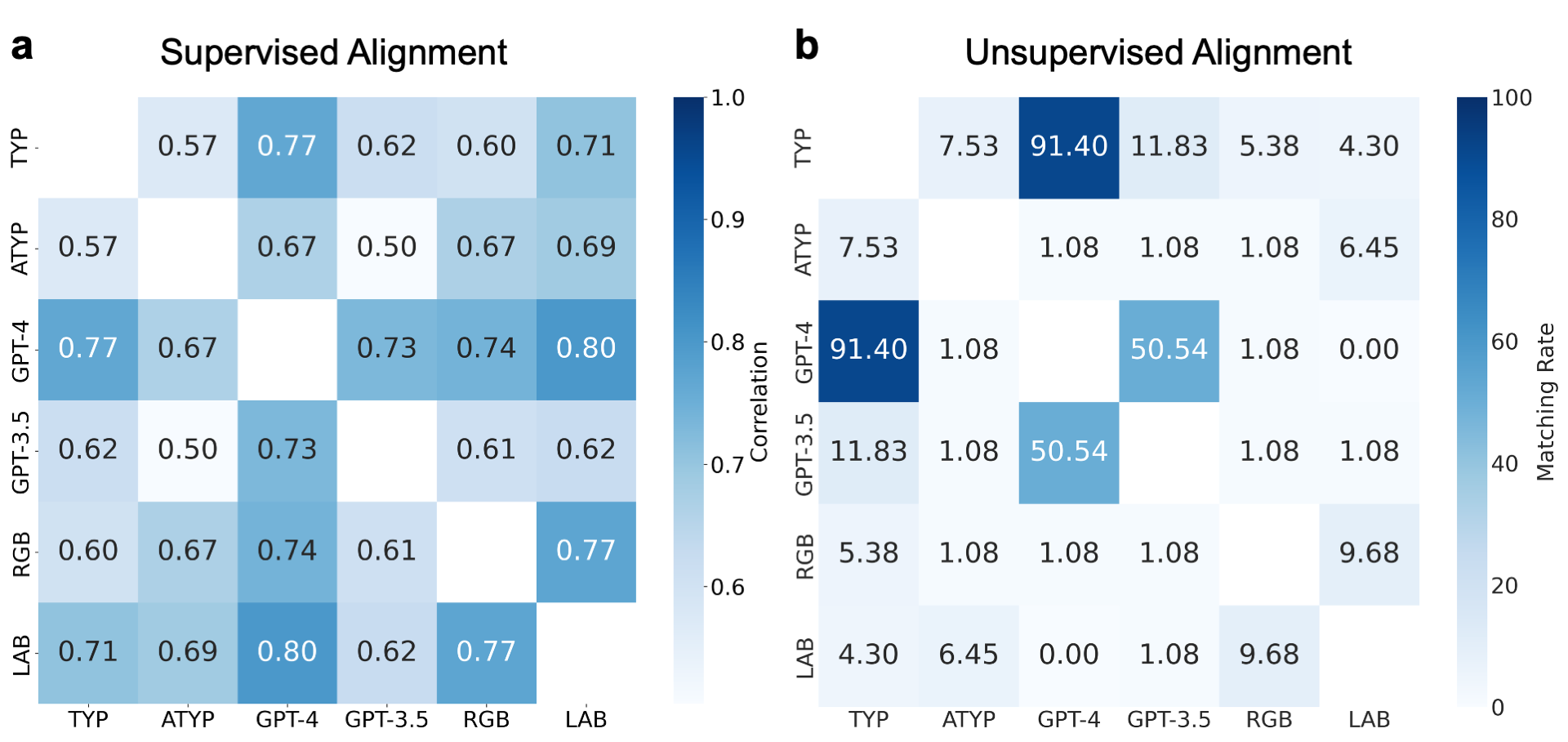}
\caption{\textbf{Evaluating the similarity of similarity structures in a supervised and unsupervised method.} (a) Conventional representational similarity analysis based on Spearman correlations. Spearman correlations between the dissimilarity matrices obtained from the color-neurotypical participants group (abbreviated by TYP), the color-atypical participants group (abbreviated by ATYP), GPT-4, GPT-3.5, and the RGB and LAB color spaces are shown. (b) Matching rates of unsupervised alignment based on GWOT between the dissimilarity matrices. 
}
\label{Fig:RSA}
\end{figure}

Before evaluating the structural correspondences of the color dissimilarity matrices between humans and LLMs as per the unsupervised alignment analysis, we first computed the similarity between them by simply computing correlations (the Spearman correlation) between them. This analysis corresponds to the conventional representational similarity analysis \cite{Kriegeskorte2013}, which implicitly assumes correspondence between the same colors across different similarity structures. Fig. \ref{Fig:RSA}a summarizes the correlations between the similarity matrices of the color-neurotypical participants group, the color-atypical participants group, GPT-4, GPT-3.5, and the RGB and LAB color spaces.

This analysis yields the following three findings:
\begin{enumerate}
    \item The dissimilarity matrix of the human color-neurotypical participant group is the closest to that of GPT-4 ($\rho=0.77$) among the models considered (GPTs and color space models). In addition, other models such as GPT-3.5 ($\rho=0.62$), the RGB color space model ($\rho=0.60$), and the LAB color space model ($\rho=0.71$) also show reasonable correlation with the similarity structure of the color-neurotypical group.
    \item GPT-3.5 shows lower correlations with the dissimilarity matrix of the human color-neurotypical group than GPT-4. In addition, the dissimilarity matrix of GPT-3.5 shows lower correlations with the color space models than GPT-4.
    \item The human color-atypical group shows relatively low correlations with the other dissimilarity matrices, suggesting that the similarity structure of the color-atypical group is significantly different from that of the the color-neurotypical group, the GPTs, and the color space models.
\end{enumerate}
The scatter plots of similarity ratings of all the pairs of the dissimilarity matrices are available in Fig. S1, providing a visual aid for understanding the correlation trends between each pair of groups.

\subsection*{Unsupervised alignment of color similarity structures}
We then evaluated the extent to which the color similarity structures of humans and LLMs could be aligned in an unsupervised manner using the Gromov-Wasserstein Optimal Transport (GWOT) algorithm. In Fig. \ref{Fig:RSA}b, we summarized the matching rates of the unsupervised GWOT alignment between all pairs of the dissimilarity matrices shown in Fig. \ref{Fig:sim}. In the following, the results of the most important pairs (color-neurotypical vs. GPT-4, GPT-3.5, LAB) are explained in more detail. The detailed results of the other pairs are shown in the Supplementary Figs. S2 and S3. 

\subsubsection*{Unsupervised alignment with GPT-4}
\begin{figure}[t]
\centering
\includegraphics[width=\textwidth]{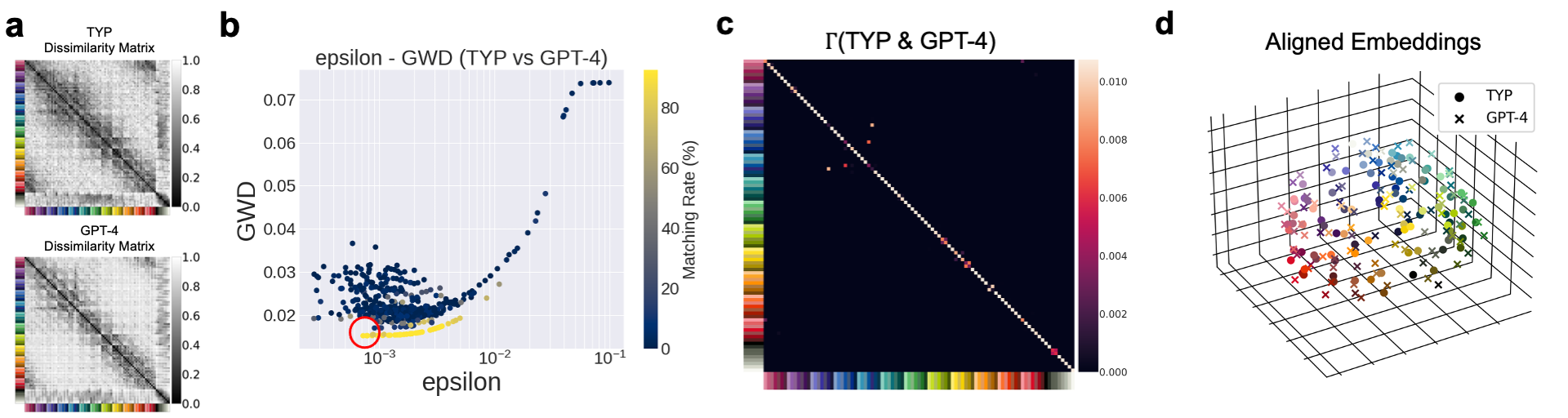}
\caption{\textbf{Unsupervised alignment between the color similarity structure of the human color-neurotypical participants and that of GPT-4}. (a) Dissimilarity matrices of 93 colors from the human color-neurotypical participants (abbreviated by TYP) and GPT-4. (b) The optimization results over 500 iterations with different $\epsilon$ values. GWD values of local minima represented by points are shown with respect to $\epsilon$. Colors represent the matching rate of unsupervised alignment. (c) Optimal transportation plan $\Gamma$ between the dissimilarity matrices of TYP and GPT-4. (d) Aligned embeddings of TYP and GPT-4 plotted in the embedded space of TYP. }
\label{Fig:GPT-4}
\end{figure}

First, we showed the results of the unsupervised alignment between the color similarity structures of the human color-neurotypical participants and GPT-4 in Fig. \ref{Fig:GPT-4}. We applied entropic GWOT to the two dissimilarity matrices shown in Fig. \ref{Fig:GPT-4}a. Since entropic GWOT is a non-convex optimization problem involving hyperparameter search of $\epsilon$, which controls the degree of entropy regularization, we performed a total of 500 optimization iterations with different $\epsilon$ values and initialization of transportation plans to search for a global optimum. The points in Fig. \ref{Fig:GPT-4}b correspond to the local minimum found in each iteration of the optimization performed on different $\epsilon$ values. Across different $\epsilon$ values, we selected the local minimum with the lowest GWD as the optimal solution (indicated by the red circle in Fig. \ref{Fig:GPT-4}b).

From the optimization process, we obtained the optimal transportation plan $\Gamma$ between the human color-neurotypical participants and GPT4 (Fig. \ref{Fig:GPT-4}c). As shown in Fig. \ref{Fig:GPT-4}c, most of the diagonal elements in $\Gamma$ have high values, indicating that most of the colors in the color-neurotypical participants correspond with a high probability to the same colors in GPT-4. To quantitatively assess the degree of correspondence, we computed the matching rate of the 93 colors (see Methods for details), which was 91.4\% (Fig. \ref{Fig:RSA}b). As can be seen in Fig. \ref{Fig:GPT-4}b, the local minima with low GWD (in the y-axis) tend to yield a high matching rate (points with yellowish color), which is necessary for unsupervised alignment to achieve a high matching rate.

To visually inspect the degree of the unsupervised alignment, we draw the embeddings of the color-neurotypical participants and the aligned embeddings of GPT-4 in Fig. \ref{Fig:GPT-4}d (See Supplementary Movies S1 for the animation of the aligned embeddings). As detailed in Methods, we aligned the embeddings of GPT-4 to those of the color-neurotypical participants by solving a Procrustes-type problem using the optimized transportation plan $\Gamma$ obtained through GWOT. Each color represents the label of a corresponding external color stimulus. Note that even though the color labels are shown in Fig. \ref{Fig:GPT-4}d, this is only for the visualization purpose and the whole alignment procedure is performed in a purely unsupervised manner without relying on the color labels. As depicted in Fig. \ref{Fig:GPT-4}d, identical colors from the color-neurotypical participants and GPT-4 are located in close proximity to each other. This shows that GPT-4 has a color similarity structure that is strikingly similar to that of the color-neurotypical participants, allowing for the successful unsupervised alignment. 

While the main results presented above were obtained using GPT-4 with text input, we also observed qualitatively similar results when using GPT-4 Vision, which takes visual input (color patches) instead of text descriptions. The unsupervised alignment between GPT-4 Vision and color-neurotypical participants revealed a high degree of structural similarity in their color representations, albeit with a slightly lower matching rate compared to GPT-4 with text input. See Supplementary Text 2 and Supplementary Figures S4 and S5 for detailed results and visualizations of the GPT-4 Vision analysis.

\subsubsection*{Unsupervised alignment with GPT-3.5}

Next, for comparison with GPT-4, we showed the results of the unsupervised alignment between the color similarity structures of the human color-neurotypical participants and GPT-3.5 in Fig. \ref{Fig:GPT-3.5} (See Supplementary Movies S1 for the animation of the aligned embeddings). The results are presented in the same format as Fig. \ref{Fig:GPT-4} and the analysis procedure is also the same as explained in the previous section. 

In contrast to the case of GPT-4, we found that the matching rate of the optimal solution (shown in the red circle in Fig. \ref{Fig:GPT-3.5}b) is much lower, 11.8\%, than that of the GPT-4, 91.4\%. However, this is still significantly higher than the chance level (1.08\%). We can also see that the optimal transportation plan $\Gamma$ in Fig. \ref{Fig:GPT-3.5}c is ``roughly'' diagonal, i.e., the diagonal elements or neighboring elements to diagonal elements of $\Gamma$ tend to have large values. This roughly diagonal appearance of $\Gamma$ means that similar colors correspond to each other between the color-neurotypical participants and GPT-3.5 with a high probability. This is also confirmed by the aligned embeddings shown in Fig. \ref{Fig:GPT-3.5}d, where the embeddings of similar colors from the color-neurotypical participants and GPT-3.5 are located close together. These results indicate that the similarity matrix of GPT-3.5, although less well aligned with the color-neurotypical similarity matrix than GPT-4, contains some structural features that can be aligned with the color-neurotypical similarity matrix.

\subsubsection*{Unsupervised alignment with color space models}

\begin{figure}[t]
\centering
\includegraphics[width=\textwidth]{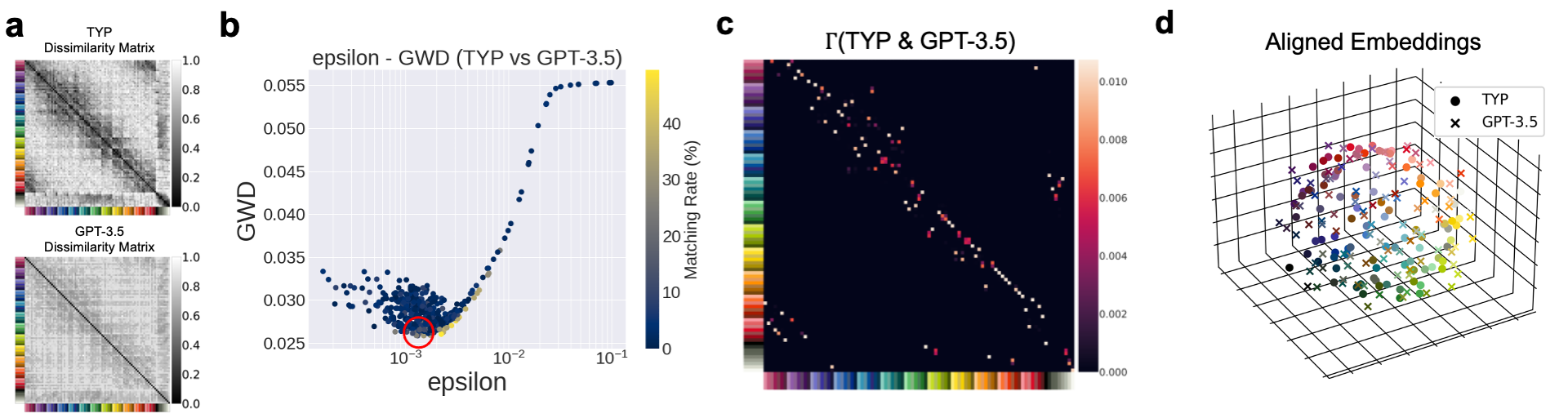}
\caption{\textbf{Unsupervised alignment between the color similarity structure of the human color-neurotypical participants and that of GPT-3.5}. (a) Dissimilarity matrices of 93 colors from the human color-neurotypical participants (abbreviated by TYP) and GPT-3.5. (b) The optimization results over 500 iterations with different $\epsilon$ values. GWD values of local minima represented by points are shown with respect to $\epsilon$. Colors represent the matching rate of unsupervised alignment. (c) Optimal transportation plan $\Gamma$ between the dissimilarity matrices of TYP and GPT-3.5. (d) Aligned embeddings of TYP and GPT-3.5 plotted in the embedded space of TYP. }
\label{Fig:GPT-3.5}
\end{figure}

To provide a baseline comparison, we showed the results of the unsupervised alignment between the color similarity structures of the human color-neurotypical participants and the LAB color space model in Fig. \ref{Fig:LAB} (See Supplementary Movies S1 for the animation of the aligned embeddings). The results are presented in the same format as Figs. \ref{Fig:GPT-4} and \ref{Fig:GPT-3.5}. 

In contrast to the both cases of GPT-4 and GPT-3.5, we found that the matching rate of the optimal solution (shown in the red circle in Fig. \ref{Fig:LAB}b) is very low, 4.30\%, which is close to the chance level (1.08\%). We also found that the appearance of the optimal transportation plan $\Gamma$ (Fig. \ref{Fig:LAB}c) is qualitatively different from those of GPT-4 (Fig. \ref{Fig:GPT-4}c) and GPT-3.5 (Fig. \ref{Fig:GPT-3.5}c). The optimal transportation plan (Fig. \ref{Fig:LAB}c) is not lined up diagonally, i.e., the diagonal elements or the neighboring elements to the diagonal elements of $\Gamma$ are small. The aligned embeddings shown in Fig. \ref{Fig:LAB}d are also quite different from those of GPT-4 (Fig. \ref{Fig:GPT-4}d) and GPT-3.5 (Fig. \ref{Fig:GPT-3.5}d), i.e., the embeddings of similar colors are not closely positioned, indicating that similar colors are not correctly aligned by the unsupervised alignment. We also obtained the similar results for the RGB color space model. The matching rate between the color-neurotypical participants and RGB is 5.38\% (Fig. \ref{Fig:RSA}b) and the optimal transportation plan $\Gamma$ does not show roughly diagonal appearance (Supplementary Fig. S2). 
\subsubsection*{Unsupervised alignment with color-atypical participants}
As another negative control, we also showed the results of the unsupervised alignment with the color-atypical participants in Supplementary Fig. S2 and Fig. \ref{Fig:RSA}b. As shown in Fig. \ref{Fig:RSA}b, the similarity structure of the color-atypical participants is not aligned with either GPT-4 or the color-neurotypical participants (the matching rate is 1.08\% and 7.53\%, respectively). Note, however, that the correlations between the color-atypical participants and GPT-4 and the color-neurotypical participants are reasonably high, $\rho=0.67$ and $\rho=0.57$, respectively (Fig. \ref{Fig:RSA}a). The subtle structural difference caused by red-green color deficiency is likely to prevent the successful unsupervised alignment between the color-atypical participants and GPT-4 and the color-neurotypical participants (see also \cite{Kawakita2023}). 

\subsection*{Comparison between conventional representational similarity analysis and unsupervised alignment}

Finally, we mention several important differences between the results of conventional representational similarity analysis and the unsupervised alignment based on GWOT. Comparing the Figs. \ref{Fig:RSA}a and b, we observe that the unsupervised alignment method was able to reveal more nuanced structural differences that were not observable using conventional representational similarity analysis. For example, the Spearman correlations between the color-neurotypical participants and the color space models ($\rho=0.60$ for RGB and $\rho=0.71$ for LAB) are reasonably high. In particular, the correlation of LAB ($\rho=0.71$) is close to the correlation of GPT-4 ($\rho=0.77$) and higher than the correlation of GPT-3 ($\rho=0.62$). However, as we showed in the previous section, the matching rates of the unsupervised alignment between the color-neurotypical participants and the color space models are very low ($5.38\%$ for RGB and $4.30\%$ for LAB),  almost at chance level. This suggests that human color similarity structures are not adequately captured by color space models, but are remarkably well captured by GPT-4 and is captured to some extent by GPT-3.5. Such nuanced structural differences between the human color similarity structure, the color space models, and GPT-4 or GPT-3.5 cannot be detected by conventional representational similarity analysis, which is based on simple correlation between similarity structures.

\section*{Discussion}
The primary objective of our work was to present a methodological advancement beyond simple correlation, brought by unsupervised alignment technique, Gromov-Wasserstein Optimal Transport (GWOT), for comparing the color similarity structures of humans and large language models (LLMs). Unlike previous studies with simple correlation analysis, our GWOT technique revealed more nuanced structural similarities and differences.

Specifically, among the models considered (GPT-4, GPT-3.5, RGB, and LAB), GPT-4 had a color similarity structure that most closely resembled that of the color-neurotypical participants with the highest matching rate with the color-neurotypical participants (91.4\%). Compared to GPT-4, GPT-3.5 is less well aligned with the color-neurotypical participants, but it still demonstrates a significant degree of alignment (11.8\%), outperforming the RGB and LAB color space models. Despite reasonably high correlation coefficients ($\rho=0.60$ for RGB, $\rho=0.71$ for LAB), somewhat surprisingly, the similarity structures of the color space models could not be aligned with that of human color-neurotypical participants in an unsupervised manner ($5.38\%$ for RGB and $4.30\%$ for LAB). These results indicate that our unsupervised alignment method can reveal nuanced structural similarities and differences between the similarity structures that are not discernible by simple correlation analysis. The qualitative difference between the GPTs and the color space models means that neither GPT-4 nor GPT-3.5 similarity judgments are the simple reflections of the RGB and LAB color space models. Rather, GPTs reflect something learned from massive textual data, or the combination of textual and visual data in the case of GPT-4.
\begin{figure}[t]
\centering
\includegraphics[width=\textwidth]{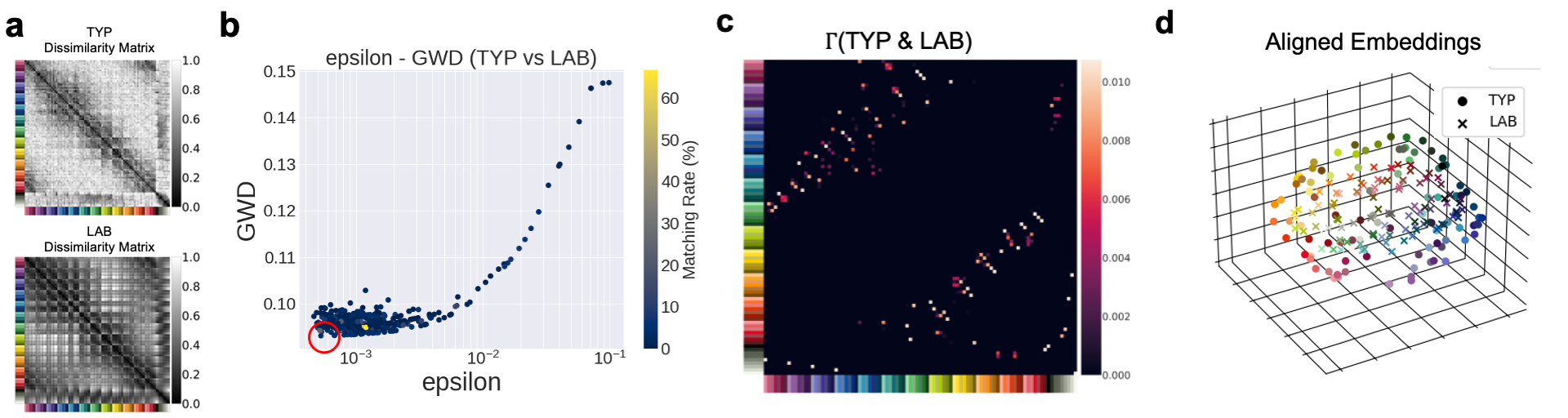}
\caption{\textbf{Unsupervised alignment between the color similarity structure of the human color-neurotypical participants and that of LAB}. (a) Dissimilarity matrices of 93 colors from the human color-neurotypical participants (abbreviated by TYP) and LAB. (b) The optimization results over 500 iterations with different $\epsilon$ values. GWD values of local minima represented by points are shown with respect to $\epsilon$. Colors represent the matching rate of unsupervised alignment. (c) Optimal transportation plan $\Gamma$ between the dissimilarity matrices of TYP and LAB. (d) Aligned embeddings of TYP and LAB plotted in the embedded space of TYP. }
\label{Fig:LAB}
\end{figure}
Despite our finding that color similarity structure of GPT-4 is remarkably well aligned with that of human color-neurotypical participants, it remains unclear whether GPT-4 maintains similar internal representations of color to humans. A valuable direction for future research would be to directly extract the embeddings of colors in GPT-4 and evaluate the similarity structures computed as distance matrices between these embeddings. However, note that the GPT-4 embeddings were not available at this time (January 2024).


For the GPT-human comparison studies including this study \cite{Marjieh2023}, it is important to consider the potential influence of cultural factors, such as language. This consideration is important even in the case of color discrimination and categorization \cite{Winawer2007}. While the exact details of GPT's training data are not publicly available, it is presumed that the model was primarily trained on English language data. Similarly, the human participants in our study were recruited from an English-speaking region. Given that both the GPT models and the human participants were from the same language/cultural background, it is possible that this shared background contributed to the strong alignment of their color similarity structures. Future research could explore how color similarity structures may differ across cultures and how the alignment between human and LLM color perception might be affected by cultural factors. Additionally, investigating the performance of LLMs trained on data from diverse cultural and linguistic backgrounds could provide insights into the role of culture in shaping color perception and categorization in both humans and AI models.

While this study focused on comparing the similarity structures of colors as an initial tractable attempt, future research could explore other sensory modalities across a broader range of tasks (e.g., visual object similarity judgment tasks \cite{Hebart2020,Hebart2023}). This could provide a more comprehensive understanding of the extent to which large language models accurately capture the similarity structures inherent in human perception. Some studies have already begun to compare the similarity structures of LLMs and humans in other domains based on simple correlation \cite{Marjieh2023, Marjieh2022a, Marjieh2022b}. Our unsupervised alignment method may provide a novel computational tool to explore more detailed structural differences or similarities between human cognition and LLMs that cannot be detected by simple correlation analysis.



\section*{Methods}


\subsection*{Collecting responses from large language models}
To obtain responses from GPT-3.5 (\texttt{gpt-3.5-turbo}) and GPT-4 (\texttt{gpt-4-0314}), we used a prompt that represented each color as a HEX code in line with a previous study \cite{Marjieh2023}.

The prompt we used is as follows:
\begin{quote}
People described pairs of colors using their hex codes. Rate the dissimilarity of the pair of colors: Color 1:[HEX code] and Color 2:[HEX code] on a scale of 0-7 with 0-1 being Very Similar, 2-3 being Similar, 4-5 being Different, and 6-7 being Very Different. Your rating should be any real number between 0 and 7. Your answer should be only the rating in the form of a number. No explanation is needed.
\end{quote}

The temperature parameter, which determines the degree of randomness in GPT responses, was set to 0.7. We ran 5 trials for each model, collecting a complete set of similarity judgment responses using the same prompt for each trial. We averaged the responses of the similarity judgments for all possible pairs of 93 colors over 5 trials and obtained the dissimilarity matrices of 93 colors from GPT-3.5 and GPT-4. 

\subsection*{Dissimilarity matrix of color space models}
To obtain the dissimilarity matrix of the RGB color space model, we computed the Euclidean distance as a measure of dissimilarity between each color pair within the 3-dimensional RGB space. For the LAB color space model, we used the CIEDE 2000 color difference formula \cite{Sharma2005} implemented in the Python package \verb|colormath (delta_e_cie2000)| as a measure of the dissimilarity between colors. Then, by computing the dissimilarities of all the pairs of 93 colors, we obtained the dissimilarity matrices of the RGB and LAB color space models.

\subsection*{Comparing Color Similarity Structures}
\subsubsection*{Conventional representational similarity analysis}
To compare the color similarity structures between humans and GPTs in a supervised manner using external color label information, we used the conventional representational similarity analysis (RSA) approach. We computed Spearman correlations between all pairs of similarity matrices (considering only the upper-triangular elements of each matrix). It is important to note that this analysis inherently assumes a correspondence between the same colors across different similarity structures.

\subsubsection*{Unsupervised alignment using Gromov-Wasserstein optimal transport}
To evaluate the similarity between the color similarity structures in an unsupervised manner, i.e., without making any assumptions about the correspondence of colors between different similarity structures, we used the Gromov-Wasserstein optimal transport (GWOT) algorithm. GWOT is an unsupervised alignment method that identifies the optimal transport plan between point clouds in two domains without requiring information about the correspondence between each item. The algorithm optimizes the Gromov-Wasserstein distance (GWD),
\begin{equation}\label{gwd}
{ \rm GWD}=\min_{\Gamma}\sum_{i,j,k,l} (D_{ij}-D_{kl}’)^2\Gamma_{ik}\Gamma_{jl},
\end{equation}
which quantifies the correspondence between the similarity structures in the two domains (Fig. \ref{unsupervised}C). In our problem setting, $D_{ij}$ denotes the dissimilarity between color $i$ and $j$ in one similarity matrix $D$, while $D'_{kl}$ denotes the dissimilarity between color $k$ and $l$ in another similarity matrix $D'$. We normalized each similarity matrix $D$ so that the values range between 0 and 1. Solving the minimization problem of GWD yields the optimal transportation plan, represented by the matrix $\Gamma^*$, which effectively aligns the color structures in the two domains in an unsupervised manner. An element of the matrix $\Gamma_{ik}$ can be interpreted as the ``probability" that the $i$-th color in one domain corresponds to the $k$-th color in the other domain.

Efficient optimization of GWD can be achieved by adding an entropy-regularization term, $H(\Gamma)$, as shown in the following equation. 
\begin{equation}\label{entropic-gwd}
{ \rm GWD}_{\epsilon}=\min_{\Gamma}\sum_{i,j,k,l} (D_{ij}-D_{kl}’)^2\Gamma_{ik}\Gamma_{jl}+\epsilon H(\Gamma).
\end{equation}
This addition has been proven to enhance optimization efficiency \cite{Peyre2016}.

The optimization problem in Eq. \ref{entropic-gwd} is non-convex, meaning that the optimal solutions found by the algorithm are local optima, with no guarantee of achieving the global optimum. To find good local minima, we conducted hyperparameter tuning on $\epsilon$ and performed random initialization of the matrix $\Gamma$ using the GWTune toolbox that we developed \cite{Sasaki2023}.  This toolbox uses Optuna \cite{Akiba2019} for hyperparameter tuning and Python Optimal Transport (POT) \cite{Flamary2021} for GWOT optimization. We used $\epsilon$ values ranging from $10^{-4}$ to $10^{-1}$ with logarithmic spacing (500 different $\epsilon$ values), in line with previous works \cite{Kawakita2023}. For each value of $\epsilon$, we used a randomly initialized matrix $\Gamma$. After finding the optimized $\Gamma$ for each $\epsilon$ value, we selected the solution that minimizes the GWD without the entropy term (Eq. \ref{gwd}) as the optimal transportation plan. 

\subsubsection*{Evaluation of unsupervised alignment}
To assess the degree of agreement between two similarity structures, we calculated the matching rate between the two dissimilarity matrices using color labels. For each color, we consider it as a match if the transportation plan assigns the highest probability between the same colors in the two similarity matrices, because the transportation plan $\Gamma_{ij}$ represents the probability or weight of transporting the $i$-th color in matrix 1 to the $j$-th color in matrix 2. More precisely, for each color $i$ from matrix 1, if $\Gamma_{ij}$ is the highest among $j\in\{1,...,n\}$ and the $i$-th color in matrix 1 and the $j$-th color in matrix 2 are the same, then we consider the $i$-th color in matrix 1 to be a match with the same color, $j$, in matrix 2. 

We denote the color labels in the two dissimilarity matrices as $c_1$ and $c_2$ respectively. The matching rate or accuracy is calculated by comparing the transportation plan $\Gamma$ with these labels. For each color $i$ in the matrix 1, denoted by $c_{1i}$, the matching condition can be formalized as:
\begin{equation}
\text{{Match}}(i) = \begin{cases} 
1, & \text{if } \Gamma_{ij} = \max_{j\in\{1,...,n\}}(\Gamma_{ij}) \text{ and } c_{1i} = c_{2j} \\
0, & \text{otherwise} 
\end{cases}
\end{equation}
This function indicates whether the $i$-th color in the matrix 1 $c_{1i}$ matches with the same color in the matrix 2 $c_{2j}$. The matching rate is then the percentage of colors in the matrix 1 that match with the same color in the matrix 2, which can be calculated as:
\begin{equation}
\text{{Matching Rate}} = \frac{{\sum_{i=1}^{n} \text{{Match}}(i)}}{n}
\end{equation}

\subsubsection*{Visualization of unsupervised alignment}
To visually assess the degree of similarity between the two color similarity structures in an unsupervised manner, we obtained 3-dimensional embeddings of 93 colors. To derive the color embeddings, we applied multidimensional scaling (MDS) to the similarity matrices, yielding 3-dimensional embeddings. We then aligned a pair of embeddings, denoted $X$ and $Y$, using the orthogonal rotation matrix $Q$. This matrix was obtained by solving a Procrustes-type problem using the optimized transportation plan $\Gamma^*$ derived from GWOT.
\begin{equation}\label{eq:procrustes-gwd}
    \min_Q \|X - QY\Gamma^* \|_F^2,
\end{equation}
where $\| \cdot \|_F$ is the Frobenius norm $\|A \|_F=\sqrt{\sum_{i,j} a_{ij}^2}$. A solution to the problem can be found through the singular value decomposition of $X (Y\Gamma^*)^{\top}$.

\bibliographystyle{unsrt}

\newpage

\subsubsection*{Author Contributions Statement}
G.K. and M.O. conceived the idea. G.K. ran experiments to collect the data from GPTs. A.Z. and N. T. designed and performed experiments to collect the data from human participants. G.K. and M.O. analyzed the data. G.K. and M.O. wrote the initial draft of the manuscript. All authors reviewed the manuscript, read and approved its final version.

\subsubsection*{Data availability}
Data for the behavioral experiments is available at \url{https://osf.io/9xwr2/}.
\subsubsection*{Code availability}
Code for the behavioral experiments is available at \url{https://osf.io/9xwr2/}. Code for the data analysis is available at \url{https://oizumi-lab.github.io/GWTune/}.

\subsubsection*{Acknowledgments}
G.K. and M.O. were supported by JST Moonshot R\&D Grant Number JPMJMS2012. N.T. and M.O. were supported by Japan Promotion Science, Grant-in-Aid for Transformative Research Areas Grant Numbers 20H05710, 23H04830 (N.T.) and 20H05712, 23H04834 (M.O.). N.T. was supported by Australian Research Council (DP180104128, DP180100396). N.T. and A.Z. were supported by National Health Medical Research Council (APP1183280) and Foundational Question Institute (FQXi-RFP-CPW-2017) and Fetzer Franklin Fund, a donor advised fund of Silicon Valley Community Foundation. We thank Dominik Kirsten-Parsch and Lonni Gomes for their help in collecting the color dissimilarity data. 

\end{document}